# Production and measurement of stellar neutron spectrum at 30 keV

Javier Praena[1], Guido Martín-Hernández[2], Javier García López[3,4]

[1] *Dpto. Física Atómica, Molecular y Nuclear, Universidad de Granada, Granada, Spain.*
[2] *Centro de Aplicaciones Tecnológicas y Desarrollo Nuclear, 5ta y 30, Playa, La Habana, Cuba.*
[3] *Dpto. Física Atómica, Molecular y Nuclear, Universidad de Sevilla, Sevilla, Spain.*
[3] *Centro Nacional de Aceleradores (Universidad de Sevilla, Junta de Andalucía, CSIC), Sevilla, Spain*

E-mail: jpraena@ugr.es

**Abstract.** In the present experiment we measure the forward-emitted angle-integrated neutron spectrum of the $^7$Li(p,n)$^7$Be reaction via time-of-flight neutron spectrometry at Joint Research Center. The proton beam is shaped to a Gaussian-like distribution with energies close to the reaction threshold upstream the thick lithium target. The angle-integrated neutron spectrum resembles a Maxwellian or stellar neutron spectrum at kT=30 keV.

## 1. INTRODUCTION

One of the most interesting topics in nuclear astrophysics is related to the nucleosynthesis of the elements beyond iron, which is mainly produced via successive neutron-capture reactions and beta decays [1,2]. The neutron capture cross section data are fundamental ingredients for the calculation of the stellar reaction rates and for reproducing the observed abundance of the elements in the Universe [3]. Since neutron velocities follow a Maxwell-Boltzmann probability distribution, the neutron capture cross section in a stellar site is known as Maxwellian-Averaged Cross Section (MACS), defined as

$$MACS \equiv \frac{\langle \sigma v \rangle}{v_T} = \frac{2}{\sqrt{\pi}} \cdot \frac{\int_0^\infty \sigma(E) \cdot E \cdot e^{\frac{-E}{kT}} dE}{\int_0^\infty E \cdot e^{\frac{-E}{kT}} dE} \quad (1)$$

Where $\langle \sigma v \rangle$ is the reaction rate per particle pair; $v_T$ is the particle most probable thermal velocity; and $\sigma(E)$ is the differential neutron capture cross section of the element involved as a function of the neutron energy. One of the most prolific sites for nucleosynthesis is the He-burning in red giants stars, where the typical temperature is 0.348 GK and corresponds to $kT$=30 keV. In consequence, the MACS at $kT$=30 keV (MACS30) are key data for stellar nucleosynthesis [4,5].

The experimental determination of the MACS can be carried out via time-of-flight (TOF) technique or via activation technique. The TOF (or differential measurement) provides the cross section as a function of the neutron energy, $\sigma(E)$, which allows calculating the MACS at several stellar temperatures; if possible, is the best option. However, if the mass sample is significantly lower than milligrams, the TOF measurement lacks of enough neutron fluence, and the MACS can only be measured by activation (or integral measurement) with high flux. Integral measurements will provide the MACS but the neutron spectrum irradiating the sample must be corrected to a stellar or maxwellian spectrum. This correction depends on the differential cross section of the isotope for which the MACS is being measuring by activation. Many isotopes lack of differential cross sections in consequence the spectrum correction cannot be properly performed in integral measurements. Thus, the best option in integral measurements is to generate an exact stellar neutron spectrum at a certain stellar energy (kT) to avoid this correction.

The production of stellar neutron beams started several decades ago. In 1966 Pönitz showed that an absolute integral cross section using proton energy near the reaction threshold for thick Li target could be determined [6]. Near the $^7$Li(p,n)$^7$Be reaction threshold, 1881 keV, the neutron beam is collimated in a forward cone and completely covered with a sample, thus, the neutron fluence could be determined with the $^7$Be activity, an absolutely measurement with no reference. Later, Beer & Käppeler showed that the angle-integrated distribution of the produced neutrons could be close to a Maxwellian at $kT$=25 keV for proton energy $E_p$=1912 keV [7]. In 1988, Ratynski & Käppeler remeasured the neutron spectrum with improved sensitivity and over a larger energy range [8]. They demonstrated that the integration over the cone of 70º of aperture yields to *quasi* Maxwellian neutron at $kT$=25 keV (QMNS-25) with a lack of neutrons above 110 keV. They showed that the MACS30 could be experimentally obtained in an integral measurement with the QMNS-25. However, small corrections due the difference between true Maxwellian at $kT$=25 keV and the QMNS-25 were needed. Then, an extrapolation from 25 to 30 keV was also performed. Since then, this method has been used for measuring MACS30 of many isotopes relevant in stellar evolution and nucleosynthesis of elements.

The correction from QMNS-25 to a Maxwellian spectrum at 30 keV (MNS-30) depends on the differential cross section; therefore, it becomes a problem for poorly known or unknown cross sections as the case of short-life isotopes. For this reason, a method to produce a Maxwellian neutron spectrum at 30 keV (MNS-30) was proposed [9,10]. With this method the correction for neutron spectrum is not needed or negligible, thus, it is adequate for integral measurements on isotopes with unknown cross sections. First measurements of MACS30 with this method were already carried out on stable isotopes [11,12]. The spectrum used in such experiments was based on the well-known angle-energy yield of the $^7$Li(p,n)$^7$Be reaction close to the threshold. Preliminary experimental confirmation of the neutron spectrum of the MNS-30 method was already carried out by means of the TOF and energy spectrum measured at forward direction with a detector a 0º [13,14]. Here, it is presented the final confirmation with the complete measurement at several angles via time-of-flight neutron spectrometry.

## 2. THE METHOD

The method for MNS-30 production is based on shaping the proton beam to a distribution that impinging onto the lithium target will produce the desired neutron spectrum [9,10]. Thus, the key point is to find the proton beam with the adequate energy distribution, so, a proper combination of proton energy and degrader or shaper foil. The shaper is determined by the thickness and material. The most convenient material is Aluminium as it was already shown in previous experiments [11,12]. For an estimation of the adequate combination SRIM code is used [15]. However, uncertainty on the Al thickness, possible inhomogeneity, or accuracy of the SRIM simulations makes the measurement of the proton distribution mandatory for this method.

Preliminary SRIM calculations showed that an adequate combination to produce a MNS-30 is a proton beam of 3.665 MeV passing through Al foil of 75-µm thickness [11]. In such conditions, a proton distribution close to a Gaussian with $E_c$=1.860 MeV and FWHM=162 keV is produced. Then, simulations based on the well-known angle-energy yield of the $^7$Li(p,n)$^7$Be reaction show that the impact onto a thick lithium target will generate a MNS-30, see [11] and references therein. Conventionally, Al foils are provided with ±10% uncertainty. A change of few micrometres in a 75-µm Al thickness produces a modification of the proton distribution and in consequence an alteration in the neutron spectrum [11]. With the direct measurement of the proton distributions all these circumstances can be ignored and the proton energy can be adjusted until the desired proton distribution is found. In fact, the final proton energy is selected on the base of the results of the measured proton distributions.

### 2.1 Measurement of proton distributions

Here, the measurement of the proton distributions produced by the crossing of the proton beam through several points of Al foil with nominal thickness of 75 µm is described. For this purpose, the Microbeam line of the 3-MV Tandem at Centro Nacional de Aceleradores (Spain) was used. This line is equipped with micrometre slits to reduce the count rate to few Hz in the existing detectors. The proton beam can impinge directly onto a Surface Barrier Silicon (SBS) detector (ORTEC) located at 0º and perpendicular respect to the beam direction. The low count rate avoids the damage of the detector and keeps its energy resolution. The setup is designed to guarantee an exactly reproducible perpendicular geometry. Figure 1 shows a picture of the chamber of the Microbeam line.

Based on the accelerator calibration, the exact energy of the proton beam was 3.665 MeV. With the direct measurement of the proton beam onto the SBS, the accelerator calibration was confirmed, thus, the terminal voltage of V=1.789 MV corresponded to 3665 keV, see Figure 2.

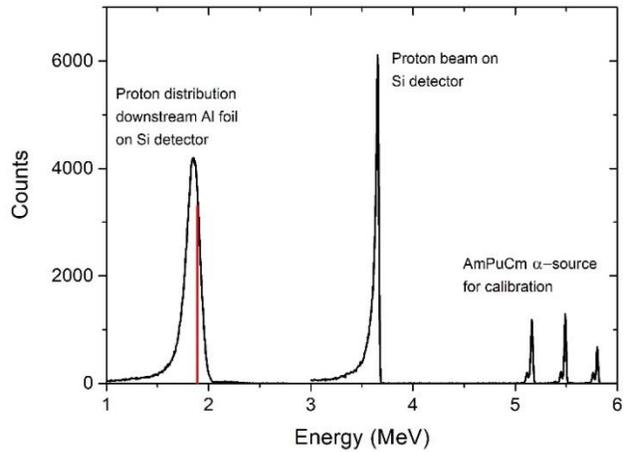

**Figure 2.** Proton distributions measured at CNA. From the right to the left: spectra and α-signals from the Am-Pu-Cm source, proton beam of 3.665 MeV impinging directly onto the Si detector (proton beam), and shaped proton beam downstream the Al foil (75-µm thickness). The spectra have the conventional long tail at lower energies due to the charge collection of the detector. Red line indicates the $^7$Li(p,n)$^7$Be reaction threshold.

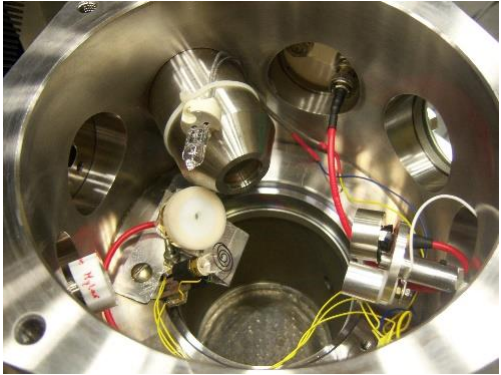

**Figure 1**. Picture of the Microbeam chamber at CNA.

Then, the Al foil was installed perpendicular to the beam and SBS detector and the proton spectrum downstream the Al foil was measured. Several points of the Al were studied to determine a possible inhomogeneity in the foil. No significant differences were found within the resolution of the SBS detector (15 keV at these energies). Figure 2 shows the measured proton distribution which can be fitted with a Gaussian distribution with $E_c$=1.860 MeV and FWHM=162 keV. The vertical red line indicates the $^7$Li(p,n)$^7$Be reaction threshold. The shape of the proton distribution monotonically decreasing with the energy is crucial to generate a stellar neutron spectrum at 30 keV.

Once the Al foil and the proton distribution were characterized, the same foil was used for the measurement of the Maxwellian neutron spectrum at 30 keV.

## 3. NEUTRON SPECTRA MEASUREMENT

The experiment for determining the neutron spectrum was performed at the 7-MV Van de Graaff laboratory at IRMM. Proton beams with energies from 1.8 to 3.7 MeV were used in the experiment for different purposes. The accelerator terminal was calibrated using the 991.86 keV $^{27}$Al(p,γ)$^{28}$Si and 2409 keV $^{24}$Mg(p,p'γ)$^{24}$Mg resonances The 90º analyzing magnet was calibrated with NMR probe. The accelerator was operated in pulsed mode with a frequency of 625 kHz for TOF measurements. The pulse width was 2-ns FWHM. Figure 3 shows a sketch of the experimental setup.

The proton beam (4-mm diameter) passed through a collimator of 5-mm diameter. The target assembly consisted of an aluminium cylinder (4.2-cm diameter and 15-cm long) and 0.4-mm thick copper backing for the lithium target. LiF thick target (6-mm diameter and 90 mg) was prepared by evaporation onto Cu backing. The target assembly was cooled by a forced air flow on the external side of the Cu target backing. Figure 4 shows a picture of the setup including target and detectors.

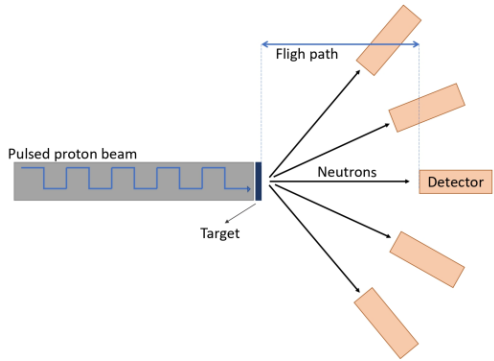

**Figure 3.** Schematic of the experimental setup. $^6$Li-glass detectors were placed on movable stands in a goniometer at the beam height. The flight path is 52 cm.

Three $^6$Li-glass detectors (5.08-cm diameter and 2.54-cm and 1.27-cm thick) detectors were located at the same height of the lithium target at a flight path of 52 cm. A fourth detector was used as monitor at a fixed position with no geometrical interference with the others. Also, a Long Counter was used as monitoring.

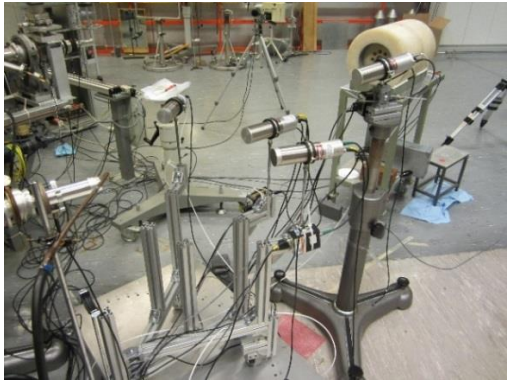

**Figure 4.** Picture of experimental setup with four $^6$Li-glass detectors, Long Counter detector and target assembly.

The acquisition was carried out from the photomultiplier of each detector; the signals were split to pulse-height and to fast-timing circuits. The latter was composed of a timing filter amplifier and a constant-fraction discriminator whereas the former consisted of a preamplifier and a spectroscopy amplifier. The timing signal from each detector provided the TOF start signal in an allocated time-to-amplitude converter (TAC). A beam pick-up monitor, placed upstream of the target, provided a signal that was processed through a timing filter amplifier and a constant fraction discriminator and provided the common TOF stop signal to the TAC. Data for every detector were acquired independently, while the timing and pulse-height information for each detector were acquired in coincidence.

## 3.1 TOF spectra without Al degrader

The experiment started checking the $^7$Li(p,n)$^7$Be reaction threshold with the measurement of TOF spectra. Then, he forward spectrum at 0° produced by 1912 keV proton energy was measured because is the standard spectrum of Ratynski & Käppeler [8]. This allows checking the acquisition setup and the analysis procedure. Details of the analysis will be described in the next section.

Figure 5 shows time spectra for three proton energies. The separation between the gammas (first peak) and the neutrons (second peaks) was very good. At proton energy of 1878 keV, neutrons were clearly detected, meanwhile no neutrons were detected at 1877 keV even the detector was located closer to the target. Therefore, an offset of 3 keV was added to the accelerator calibration. In a first calculation at 1890 keV, the maximum neutron energy corresponds to around 73 keV (141 ns), in good agreement with kinematics of the $^7$Li(p,n)$^7$Be reaction. The shape of the gamma-flash peak is analogous to a similar measurement performed in the same facility at 1912 keV [16].

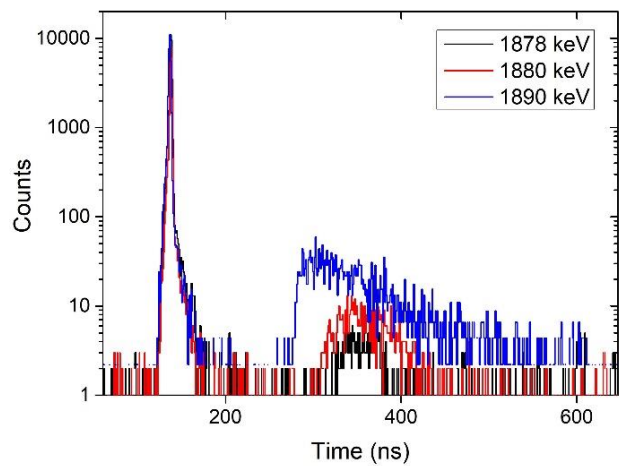

**Figure 5.** Time histograms for the $^6$Li-glass detector at 0° for three proton energies close the threshold. First peak corresponds to prompt gammas or gamma-flash that is used as time zero. Second peaks correspond to the neutrons. Channels have been already transformed to time.

Hereafter, the 3 keV offset is already considered. Then, the accelerator was set at 1912 keV proton energy and the spectrum at 0º was acquired. As mentioned before, the angle-integrated spectrum at 1912 keV proton energy has been measured several times [16-18] and corresponds to the standard neutron field established by Ratynski & Käppeler in 1988. This spectrum has been extensible used in MACS measurements where the neutron spectrum at 0º was always acquired and analysed as a further verification of the conditions of the experiments regarding proton energy and neutron production. Figure 6 shows a 0º time histogram. First analysis shows that the maximum energy of neutrons will be around 112 keV (113 ns) in good agreement with the expected result. However, the shape of the gamma-flash suffers of an increase in the FWHM compared to lower proton energies. This will be a key point of the analysis of the neutron spectra at higher proton energies with the Al foil, as it will be explained later.

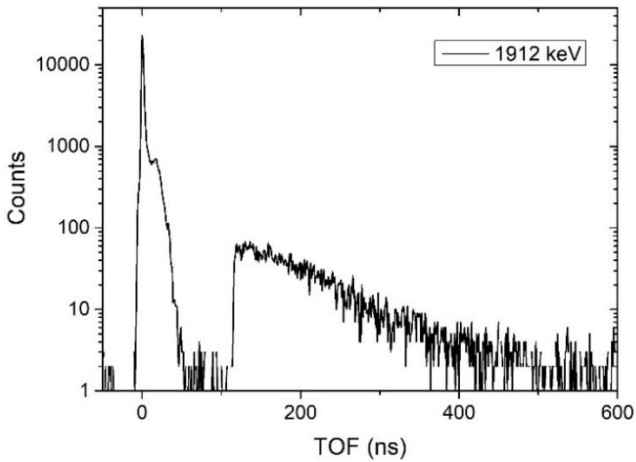

**Figure 6.** Time histogram for the $^6$Li-glass detector at 0º for 1912-keV proton energy.

## 4. DATA ANALISIS

The data analysis of the TOF spectra is determined by the time-to-energy conversion of the neutrons. First, it depends on the $^6$Li-glass detector efficiency, mainly determined by the $^6$Li(n,t)$^4$He reaction. Some resonances in other materials of the detector have been already studied. Lederer *et al.* [17] and Macías *et al.* [18] with similar $^6$Li-glass detectors showed a small contribution to the intrinsic efficiency of resonances of $^{56}$Fe and $^{28}$Si which are also present in the Borosilicate, µ-metal and tape of the detectors. Using the reported geometry and materials of the $^6$Li-glass detectors, the intrinsic efficiency is obtained with Monte Carlo simulations with MCNPX [19]. Figure 7 shows the simulated intrinsic efficiency. Small dips due to the mentioned resonances can be noticed.

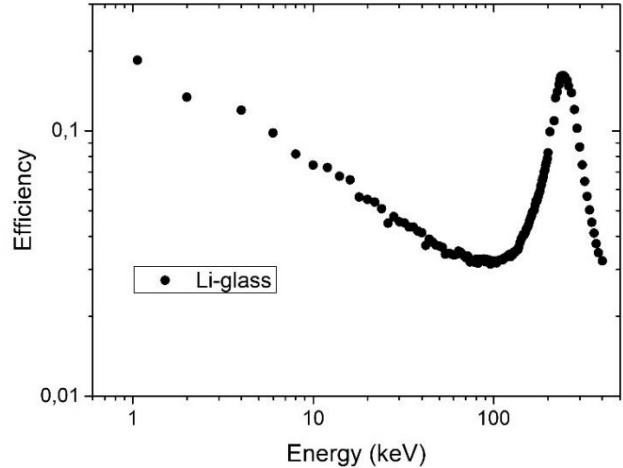

**Figure 7.** Intrinsic efficiency of a $^6$Li-glass detector (5.08-cm diameter and 2.54-cm thick).

Second, the final TOF-to-energy conversion depends on the geometry of the setup. For this, all the processes until a neutron induces the $^6$Li(n,t)$^4$He reaction inside the detector must be considered. These processes are an effective increment of the flight path of the neutrons inside the detector which depends on the energy of the neutron. The TOF-to-energy conversion was obtained with accurate simulations with MCNPX considering the whole setup.

The good separation of gammas and neutrons in Figure 5 and 6 shows the presence of some scattered signals between both peaks. This background must be subtracted. A run with a shadow bar shielding the detector showed that scattered signals are due to backscattered neutrons in the experimental hall. However, this background was very small and uncorrelated with the time (flat background) and can be subtracted.

The uncertainty of the neutron spectrum at 1912 keV at 0º is mainly due to statistics. Therefore, once the time-of-flight is transformed to an energy grid, the histograms are rebinned to 5 keV in the region from 2 to 120 keV. Figure 8 shows the comparison between the experimental neutron spectrum at 0º of this work with 1912 keV proton energy and the experimental result of Ratynski & Käppeler [8,17]. The agreement is very good and demonstrates the feasibility of the setup, acquisition system and analysis procedure. The same type of analysis has been successfully used in previous works [18,20].

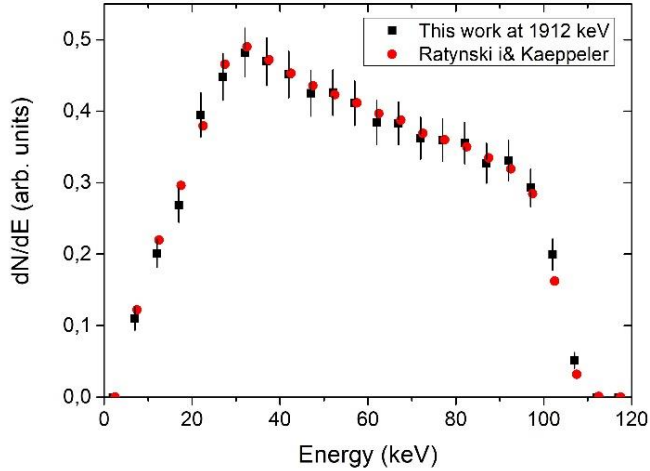

**Figure 8.** Experimental neutron spectrum of this work at 0º and 1912-keV proton energy in comparison to the experimental data of Ratynski and Käppeler [8,17].

### 4.1 TOF spectra with Al degrader

Once the measurements at proton energies near the $^7$Li(p,n)$^7$Be reaction threshold were performed, the characterized Aluminum foil was placed close to fresh LiF target. Figure 9 shows a picture of the target assembly during the dismounting, where the Al foil is perpendicular to the target by a pushing system made of carbon fiber and located inside the target assembly. At the end of the experiment, a mark was clearly visible in the central part of the Al foil. This mark was made of sputtered Lithium due to the impact of the proton beam on the target. Even the low average intensity delivered by the pulse proton beam (around 100 ns in continuous mode) a loss of $^7$Be was found. This matter in absolute integral measurements as the fluence is determined with the $^7$Be activity. The Lithium target is deposited onto the Cu backing and it became dark by proton irradiation as conventionally occurs in experiments with LiF. The Cu backing acted as proton beam dump and close of the vacuum of the accelerator by suction and its contact with a Viton O-ring.

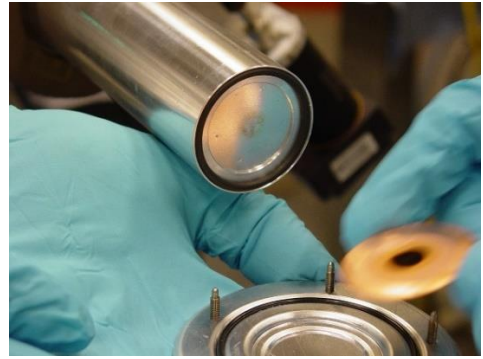

**Figure 9.** Picture of the target assembly. Aluminium foil has a mark where the proton beam passed through. This mark was due to sputtered Lithium, thus, it contained $^7$Be. LiF is deposited onto the Cu backing and become dark by proton irradiation.

Then, the accelerator was adjusted to 3.665 keV proton energy and time histograms were acquired. The protons passed through the Al foil before impacting the LiF target. Figure 10 shows a time histogram for the $^6$Li-glass detector at 0º.

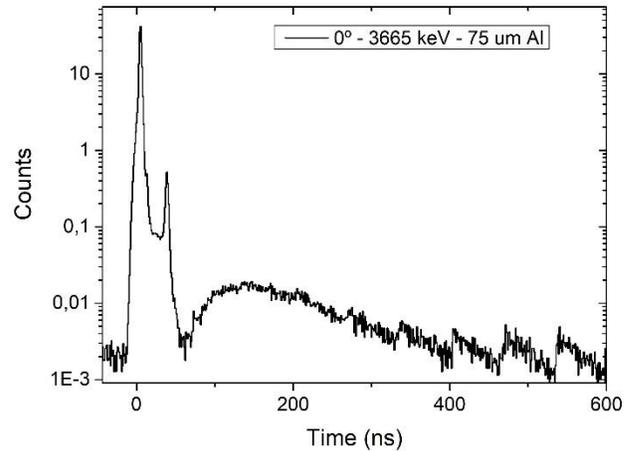

**Figure 10.** Time-of-flight histogram recorded by the $^6$Li-glass detector at 0º for 3665-keV proton beam passing through the 75-µm Al foil and hitting the thick LiF target. Counts are normalized to the monitoring detector.

Important differences with the time spectra at lower proton energies appeared. First, there is not a clear separation between gamma-flash and neutron peaks. This is an indication of a background correlated with the time. Second, a clear structure of double peak appears in the gamma-flash, and it was also visible at higher times (above 300 ns). Therefore, the experiment deals with an important background that must be understood and properly subtracted. After several test with the chopper and buncher system, the problem was found out. It was related to the presence of a halo in the proton beam at the ion source of the accelerator. This halo increased with the proton energy. Thus, protons of the beam halo passed through the collimator connected to the chopper system even the main part of the proton beam was not passing through the collimator. Then, the buncher provides the final proton beam time structure. The structure of the halo and the time structure provided the second large gamma-flash and the contamination at whatever time.

Although the background in the experiment was important, it can be subtracted once the problem has been understood. The background subtraction is performed by the subtraction of the second gamma-flash and the corresponding neutrons, and the repetitive structure. Figure 11 shows this procedure. The repetitive structure was acquired at angles higher than 80º were neutron fluence is negligible. The result of the substruction is shown in Figure 12.

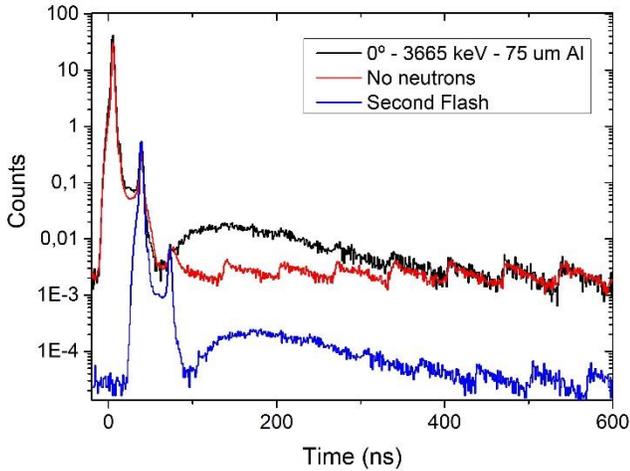

**Figure 11.** Time-of-flight histogram recorded by the $^6$Li-glass detector at 0º (black), and the two histograms to be subtracted, at angle higher than 80º (red), and the 2$^{nd}$ γ-flash and neutrons (blue). Counts are normalized to the monitoring detector.

A preliminary analysis of the TOF histogram at 0º after the background subtraction shows that the shape of the neutron peak at lower TOF is monotonically increasing with the time instead of a sharp shape as in the case of the TOF histogram at 0º acquire at 1912-keV proton energy. This is an indication of a longer tail at high neutron energies as expected for the MNS-30. Besides, the highest neutron energies will be around 250 keV (75 ns). It should be remembered that the Ratynski & Käppeler angle-integrated spectrum has a lack of neutrons above 110 keV.

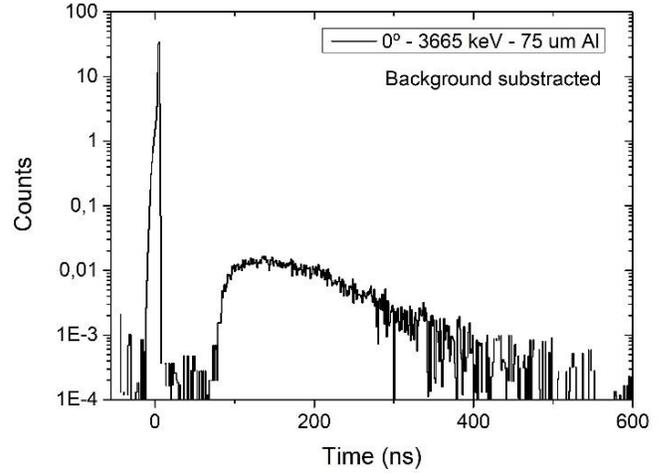

**Figure 12.** Time-of-flight histogram at 0º after the background subtraction explained in the text.

The same subtraction procedure is applied to the rest of the measured angles, from 0º to 80º in steps of 10º. Following the TOF-to-energy conversion already explained for 1912 keV, the energy histograms are obtained for each detector and angle. Figure 13 shows the neutron spectra for each detector normalized to the number of counts registered in the monitoring detector. The results are shown with a constant bin of 5 keV which is a good compromise between good energy resolution and reasonable statistics. With this binning the statistics uncertainty ranges from 3 to 12%. The uncertainty is not included for display purposes.

Once the neutron spectra are determined at detector position, it is needed to obtain the angle integrated neutron spectrum at the source point, thus, at the neutron production target. For this, each spectrum is scaled by the respective covered solid angle. For angles (α) higher than 0º the corresponding factors are calculated as,

$$f_\alpha = \cos(\alpha - 2.79°) - \cos(\alpha + 2.79°)$$

And for the 0º position as,

$$f_0 = 1 - \cos(2.79°)$$

Where 2.79º is the half of the covered angle by the detector. Figure 14 shows the neutron spectra for each detector normalized to the number of counts registered in the monitoring detector and corrected by the corresponding solid angle.

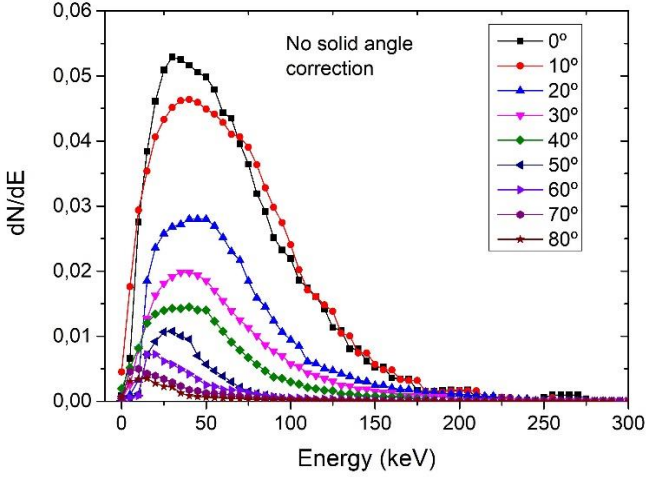

**Figure 13.** Neutron spectra from 0º to 80º in steps of 10º. Spectra are not corrected for the respective solid angle.

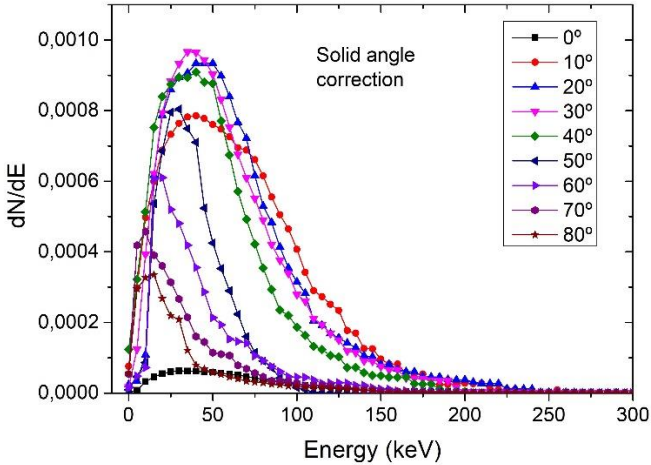

**Figure 14.** Neutron spectra from 0º to 80º in steps of 10º. Spectra are corrected for the respective solid angle.

For the present setup with a flight path of 52 cm and 5.08-cm diameter detectors, the whole forward hemisphere is not covered. However, the possible differences of the angle-integrated spectrum have been studied for other authors at 1912-keV proton energy [17]. They demonstrated that the possible differences are very low. Figure 15 shows the final angle-integrated experimental spectrum obtained in the present work (black histogram). The experimental data are fitted with a Maxwell-Boltzmann distribution in energies (red line),

$$f_{MB} = A \cdot E \cdot e^{-E/kT}$$

with the normalization constant (A) and the stellar energy (kT) as free parameters. With an orthogonal linear regression is obtained $kT=30\pm0.3$ keV with a reduced $\chi^2=5.4\cdot10^{-6}$ and $R^2=0.99$. The results show a very good agreement between the experimental neutron spectrum and a stellar neutron spectrum at $kT=30$ keV.

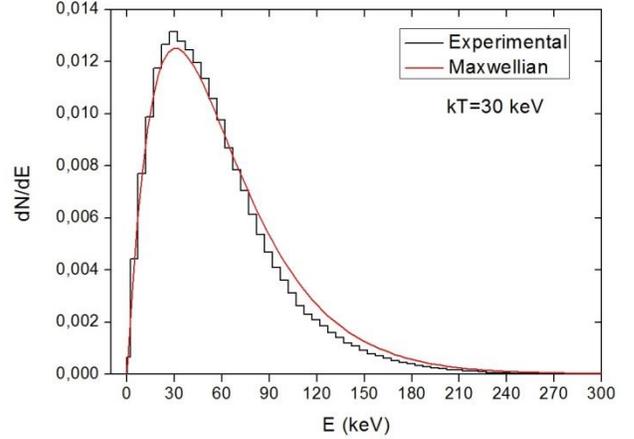

**Figure 15.** Experimental neutron spectrum generated in the present work (black), Maxwell-Boltzmann fit (red) with free parameters the normalization constant and the stellar energy. The corresponding result is kT=30±0.3 keV.

## 5. CONCLUSIONS

The main purpose of the present work was the production and measurement of a stellar neutron spectrum at $kT=30$ keV. The goal was to improve the standard neutron field produced by Ratynski and Käppeler in 1988. Since then, such spectrum has been used in several integral measurements of MACS30 of important isotopes for astrophysics. The improvement was focused on the possibility to avoid the spectrum correction for determining the MACS. The spectrum correction depends on the differential cross section of the element for which the MACS is determined by an integral measurement. Therefore, for elements with poor or unknown knowledge of the differential cross section, this correction becomes a puzzle. The generation of a Maxwellian neutron spectrum at 30 keV solves this problem. In order to improve the Ratynski and Käppeler spectrum, it was needed that the Maxwellian fit of the neutron spectrum was increased from 25 keV to 30 keV, and at the same time to generate neutrons with energies above 110 keV following a tail at high energies. Both goals have been achieved with the use of a shaped proton beam onto a thick lithium target.

It is important to mention that the energy distribution of the proton beam must be monotonically decreasing from the threshold of the reaction to around 2.1 MeV. As it has been shown, the direct measurement of the proton beam is possible and it is a key of this method. Then, the measurement of the stellar neutron spectrum at 30 keV has been carried out via time-of-flight technique. Even, the important background suffered in the experiment, it is possible to subtract it from each spectrum acquired at different angles. The result of the angle-integrated spectrum resembles a stellar or Maxwellian at 30 keV. Very small differences are noticed in the peak of the Maxwellian and above 100 keV. However, the improvement of the stellar shape compared to Ratynski and Käppeler is noticeable. Thus, the goal of the experiment has been achieved. Additionally, the MNS-30 method solves a technical problem which is the loss of $^7$Be in absolute measurements. As it has been shown, the sputtering of Lithium occurs even at low currents but the Al foil avoids its loss to the accelerator.

After the experimental confirmation of the stellar spectrum at 30 keV, it is planned to carry out an absolute integral measurement of the MACS30 of $^{197}$Au(n,γ). This quantity is extremely important in stellar nucleosynthesis and its value is matter of discussions nowadays due to the discrepancy between the integral measurement of R&K and the value obtained with differential measurements of the $^{197}$Au(n,γ) cross section [21-26]. In a forthcoming paper it will be demonstrated that the neutron correction related to the MNS-30 is not needed because the folding of the $^{197}$Au(n,γ) cross section with a Maxwellian at 30 keV is practically equal to the folding with the MNS-30 measured in the present work.

**Acknowledgments.** This work was supported by the European Commission within the Seventh Framework Programme through EUFRAT (EURATOM contract no. FP7-211499). The authors would like to the operators of the IRMM Van de Graaff accelerator, in particular to Thierry Gamboni, for the many extra hours they devoted trying to provide the best possible conditions for this experiment. The authors thank J.A. Labrador and A. Romero for the excellent performance of the CNA accelerator. J.P. acknowledges support and constructive discussions with F. Käppeler in the framework of the n_TOF-CERN Collaboration. This work was also supported by Spanish institutions Ministerio de Ciencia e Innovación (PID2020-117969RB-I00), Junta de Andalucia projects P20-00665 and B-FQM-156-UGR20.